\documentclass[conference]{IEEEtran}
\IEEEoverridecommandlockouts
\usepackage{cite}
\usepackage{amsmath,amssymb,amsfonts}
\usepackage{algorithmic}
\usepackage{graphicx}
\usepackage{textcomp}
\usepackage{xcolor}
\usepackage{lipsum}
\usepackage{graphicx}

\def\BibTeX{{\rm B\kern-.05em{\sc i\kern-.025em b}\kern-.08em
    T\kern-.1667em\lower.7ex\hbox{E}\kern-.125emX}}
\begin{document}

\title{Scalable data storage for PV monitoring systems
}

\author{\IEEEauthorblockN{1\textsuperscript{st} Anastasios Kladas}
\IEEEauthorblockA{\textit{Faculty of Technology Engineering} \\
\textit{KU Leuven}\\
Ghent, Belgium \\
Anastasios.Kladas@kuleuven.be}
\and
\IEEEauthorblockN{2\textsuperscript{nd} Bert Herteleer}
\IEEEauthorblockA{\textit{Faculty of Technology Engineering} \\
\textit{KU Leuven}\\
Ghent, Belgium \\
Bert.Herteleer@kuleuven.be}
\and
\IEEEauthorblockN{3\textsuperscript{rd} Jan Cappelle}
\IEEEauthorblockA{\textit{Faculty of Technology Engineering} \\
\textit{KU Leuven}\\
Ghent, Belgium \\
Jan.Cappelle@kuleuven.be}
}

\maketitle

\begin{abstract}
Efficient PV research which includes a prolonged data monitoring from multiple experiments with different characteristics, requires a scalable supporting system to handle all of the collected information. This paper presents the development of a relational database for hosting all the necessary information for data modeling, comparative analysis and O\&M systems. Ramer-Douglas-Peucker algorithm and Timescaledb compression are used to decrease the size of the time-series data and increase the performance of the queries. A decision-making algorithm is presented for selecting the optimal inputs to the Ramer-Douglas-Peucker algorithm to ensure the maximum disk space savings while not losing any of the necessary information. Furthermore, alternative ways of implementing the same database are provided.
\end{abstract}


\section{Introduction}
The urgent need for clean energy production has increased the PV research \cite{ELAMIM2018121}. Research institutes, organizations, or companies often need to monitor several measurement sites with different characteristics (PV modules, climate measurements, etc.) and make conclusions through comparative analysis. On the other hand, as the data science has started to be involved in PV research \cite{Arafet,Naik}, there is a need to store efficiently high-resolution historical data for training efficient data-driven models. 

A common approach presented in PV literature is the data storage into log files locally on a dedicated PC \cite{4Mujumdar,5Touati,6GAD2015337,7HERTELEER2017408}. This method despite its simplicity requires a big manual effort from the user and a lot of processing power (loading all data before filtering etc.), increasing the hardware requirements especially in cases of long period analyses. 
Implementations of relational databases can be found in the literature \cite{8Zdanowicz,9Meliones,10Nihar}. The issue with those approaches is that their design is motivated by a single research objective. Therefore they will need significant modifications on further additions (different experimental characteristics, additional measurements, adjustable sampling frequency, etc.).

The data storage usually takes place in tables or datasheets containing multiple columns for each of the measurements. Despite its query simplicity, this approach requires a fixed sampling resolution for all readings even if the rate of change between the data points differs among them. Thus, extra disk space will be needed to store all information that could be avoided. Taking the above into consideration, an accessible centralized database approach would be beneficial.

Current work presents a scalable and powerful relational database structure for PV research. The relations between the tables are heeding the physical connections between modules-equipment and the objective is the ability to store high definition historical and current measurement data while employing the minimum amount of disk space, to be fast and not need any major structural changes on additions. This database is optimized for simulations, comparative analysis, modeling, and monitoring purposes but it can be used in a broader range of applications.

\section{Database}
The whole database is developed using PostgreSQL 13. It runs on a HP workstation and it is accessible from the local network or remotely after VPN connection. 
\subsection{Structure}

\begin{figure*}
    \centering
    \includegraphics[width = \linewidth]{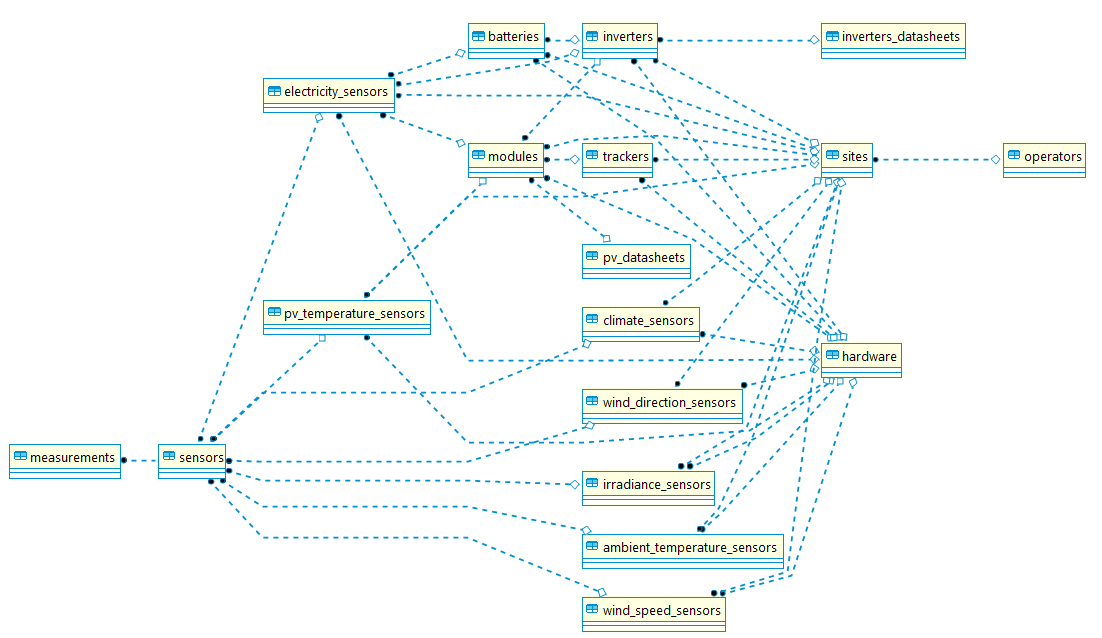}
    \caption{Graph representation of the examined schema generated by DBeaver software.}
    \label{fig:schema}
\end{figure*}

The database starts with the operators of every site and ends with the measurement of every sensor (fig. \ref{fig:schema} right to left). In other words, every measurement is connected with all information regarding its related hardware and people in charge. A common approach to storing measurement data is the usage of NoSQL time-series databases for IoT applications as they lean to be faster and lighter \cite{10Nihar, 11Percuku}. In this approach, the measurements table which hosts all measurements is implemented as a time-series table using the extension of PostgreSQL, Timescale DB. In this way, a combination takes place between the advantages of relational database (robustness and reliability) and the highest performance of NoSQL, while filtering and manipulating the data using the same SQL queries. 

The graphical representation of the presented schema is displayed in fig. \ref{fig:schema} and the utilization of all tables is described in the table \ref{tab:tables}.

\begin{table*}[ht]

\centering
\caption{Brief explanation of every table row and its dependencies.}
\label{tab:tables}
\begin{tabular}{p{0.2\linewidth}  p{0.5\linewidth} p{0.3\linewidth}}

\textbf{Table name}                  & \textbf{Purpose}                                                                                                                                                                                                                                                             & \textbf{Connected with}                                            \\

\hline
Operators                   & Information about the people in charge of every experimental site.                                                                                                                                                                                                  &                                                                        \\

\hline
Sites                       & Site information :  latitude, longitude, elevation, etc.                                                                                                                                                                                                            & Operators                                                             \\
\hline
Hardware                    & Serial numbers of every device/module that is used                                                                                                                                                                                                                  &                                                                       \\
\hline
Trackers                    & The PV trackers and their characteristics.                                                                                                                                                                                                                          & Sites and hardware                                                     \\
\hline
Inverter datasheets         & Inverter characteristics                                                                                                                                                                                                                                            &                                                                       \\
\hline
Inverters                   & The inverters used.                                                                                                                                                                                                                                                 & Sites, hardware and inverted datasheets                            \\
\hline
Batteries                   & Information about the batteries installed.                                                                                                                                                                                                                          & Inverters, sites and hardware                                         \\
\hline
PV datasheets               & PV specifications                                                                                                                                                                                                                                                   &                                                                       \\
\hline
Modules                     & PV modules used and installation characteristics (Orientation(s), tilt angle(s), etc.).                                                                                                                                                                             & PV datasheets, inverters, sites, trackers and hardware               \\
\hline
Electricity sensors         & All electricity related sensors (voltage, current, power).                                                                                                                                                                                                          & Batteries, inverters, sites(for easier queries), modules and hardware  \\
\hline
PV temperature sensors      & Information about the PV temperature sensors                                                                                                                                                                                                                        & Modules, sites and hardware                                          \\
\hline
Irradiance sensors          & Information about the irradiance sensors’ installation characteristics (tilt, orientation).                                                                                                                                                                         & Sites and hardware                                                     \\
\hline
Ambient temperature sensors & Information about the ambient temperature sensors.                                                                                                                                                                                                                  & Sites and hardware                                                      \\
\hline
Wind speed sensors          & Information about the wind speed sensors.                                                                                                                                                                                                                           & Sites and hardware                                                     \\
\hline
Wind direction sensors      & Information about the wind direction sensors.                                                                                                                                                                                                                       & Sites and hardware                                                     \\
\hline
Climate sensors             & Information about the less frequent climate sensors (PAR, humidity etc.).                                                                                                                                                                                           & Sites and hardware                                                     \\
\hline
Sensors                     & Table were all IDs of sensor tables are concentrated and take a new ID for the measurements table (as "Sensor ID"). Trigger functions have been build to the above sensor tables for automatic inserting their new IDs to this table on every new sensor entry. & All sensor tables                                                      \\
\hline
Measurements                & Table were all measurements are stored.                                                                                                                                                                                                                             & Sensors                                                               
\end{tabular}
\end{table*}

\section{Compression}
Dedicating one row for each high-resolution measurement (e.g. every second), results in a table of tremendous size. To face this issue, a two-step compression is taking place on the measurements dedicated table.

\subsection{Ramer-Douglas-Peucker algorithm}\label{AA}
The purpose of this step is to compress the data in such a way that they can be retrievable via linear interpolation when there is a need of working with them. For that reason the Ramer-Douglas-Peucker algorithm \cite{12RAMER1972244, 13Douglas1973ALGORITHMSFT} will be used.  This algorithm is used to decrease the data points of a polyline while preserving its characteristics/shape. Based on the Euclidian distance between the points and a factor symbolized by the Greek letter epsilon ($\epsilon$), it dismisses the data points that consist (or they are very close) to a line, keeping its first and the last point. The compression size is proportional to the epsilon value.

The data could be transmitted to the server in several ways (MQTT, HTTP, Cloud). If the data is delivered in a file format, the algorithm is performed to the entire file, as soon as the file arrives. 

Otherwise, on live-streamed data, all data are stored in temporary tables in a file-based local database (SQLite). On the current system, the sampling period of the live streaming data is one second. An algorithm runs periodically (on the current running system every five minutes) to move the data from the file-based database to the main one, after compressing them in a separate thread (to not interrupt the data collection procedure). The algorithm runs separately for every different sensor. More specifically, when the algorithm is executed, it stores the first and the last timestamps from the related table (on the file-based database) for the examined sensor and loads all the data in between them. It resamples them to one second periods and performs linear interpolation to fill missing data points between the samples (if they exist). Then the Ramer-Douglas-Peucker algorithm is performed to reduce the data points. Afterward, the new compressed dataset is appended to the database. Meanwhile, the timestamp of the execution is stored to calculate the time until the next one. Then the data between the first and the last timestamp are deleted from the source temporary table.

Epsilon is determined by taking into consideration the value range of the measurement and the fluctuation of the respective sensors. A practical example is the following. 

\begin{figure*}
    \centering
    \includegraphics[width = \linewidth]{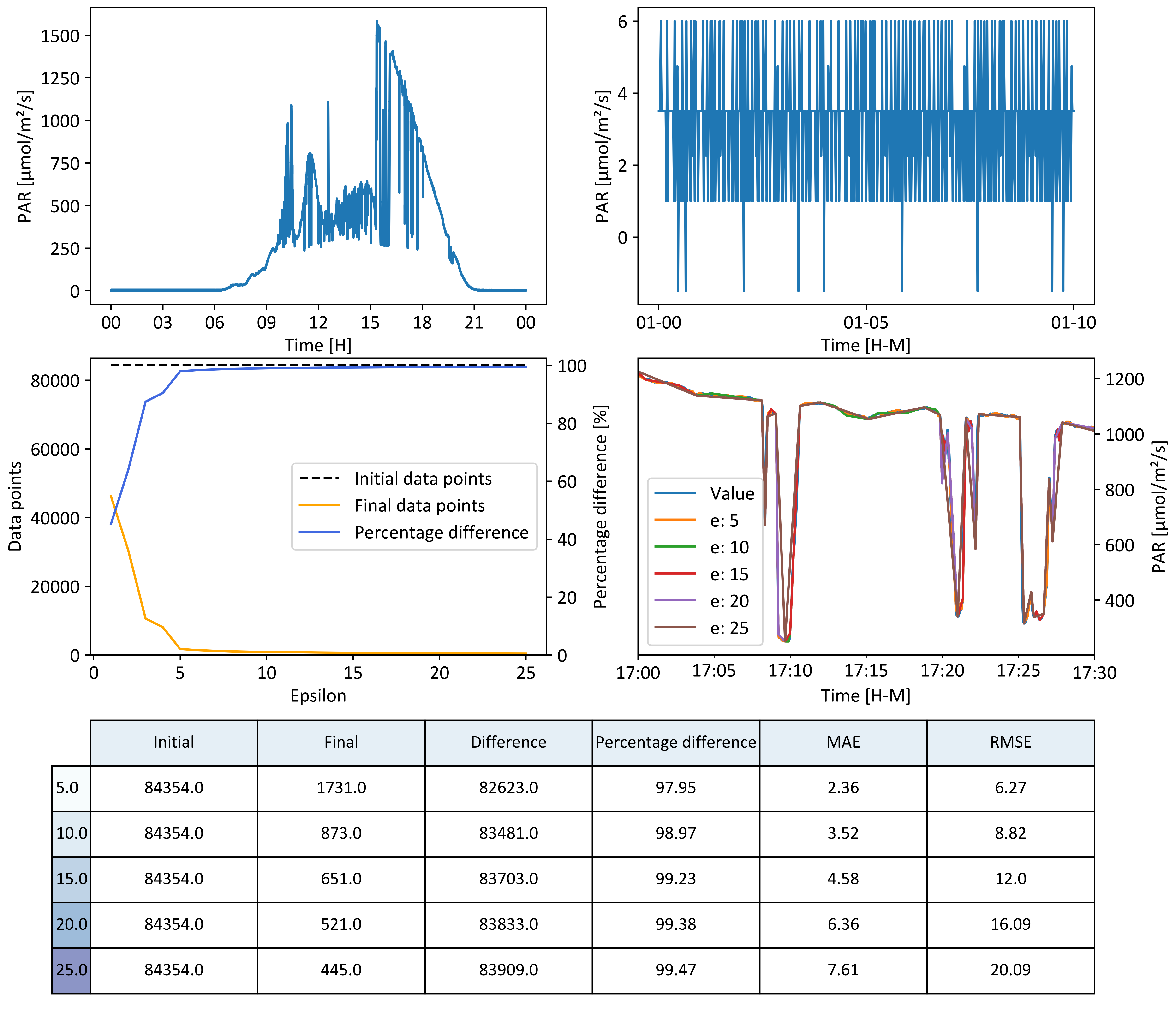}
    \caption{Graphical representation of the procedure to determine the optimal epsilon for each PAR lighting. Upper left: Selection of a day with high fluctuation. Upper right: Visualization of the sensor fluctuation or noise at night). Middle left: Difference between initial data points and final after the compression with various epsilon. Middle right: Visualization of the initial along with the compressed signals on a time interval with high fluctuations. Bottom: Table comparing the initial and the final (after compression and interpolation) data-points.}
    \label{fig:finding epsilon}
\end{figure*}

In this example, the decision-making procedure takes place for PAR light measurements. The data has been assessed from Apogee Model SP-214 PAR sensors, from the rooftop of the tallest building of KU Leuven, Ghent. A high fluctuation day (upper left fig. \ref{fig:finding epsilon}) has been selected from an old database with one second sampling period, based on the maximum daily average value of the standard deviation for every two hours interval. For the noise detection of the sensor, an observation takes place on its readings during steady state conditions. As the examined sensor works with irradiance, the observation takes place at night time where zero values are expected. As it can be noticed from fig. \ref{fig:finding epsilon} (upper right plot),  the reading fluctuate between one and six. Therefore an epsilon equal to five is expected to result in minimal/zero exploitable information loss. As several epsilon values have been tested (middle left fig. \ref{fig:finding epsilon}), the results show that there is a tremendous data point decrease comparing to the initial signal. More specifically, for epsilon equal to five, 98\% less disk space is required, while for epsilon equal to 25, 99.5\%. On the time-series comparison line plots between the signal and the compression (middle right fig. \ref{fig:finding epsilon}), it can be seen that the deviations from the initial signal start to be noticeable for epsilon higher than 10. Therefore, considering that for epsilon values higher than five the storage savings are similar while the error metrics are increasing almost linearly (bottom table fig. \ref{fig:finding epsilon}), epsilon has been selected to be equal to five.

Following the same approach, epsilon values are selected for every sensor category.

The fewer remaining data points lead also to higher query performance on the database layer. 

\subsection{Timescale DB compression}
Even more disk space can be saved if the compression built-in function of Timescale DB will be used. More specifically, when compression is enabled, it converts the rows of stored data into arrays containing the key-value pairs between date-time and measurements, for every partition of the table (which is based on the sensor identification key), ending up with fewer rows.  The compression policy of the compression function determines the size of the data stored in each array. In this work the selected compression policy is one day, meaning that all data from each day are concentrated in one row.   Apart from the saved disk space, compression can also speed up some queries.

\section{Discussion}
The proposed data structure is optimized concerning scalability and the ability to support a wide range of applications minimizing the manual work of the developer or analyst. The motivation behind the database is the implementation of O\&M systems able to support the optimized PV performance as well as to send notifications on unexpected circumstances. Also, it supports advanced dashboards.

The same schema could be implemented using one single table for the sensor characteristics where a column with JSON data type would be used to store all special sensor type characteristics. In that manner, queries of measurements would become much simpler, and also build trigger functions (for inserting the IDs to the sensors table on every new sensor insert) could be avoided. On the other hand, accessing exceptional information for every sensor would become more complicated. For sake of simplicity, manufacturer characteristics for every sensor are included in the sensor table and not as PV modules and inverters which have special tables for their datasheets. This results in fewer tables, but repeatable information on certain cases. Furthermore, ADC converters and controllers have been neglected from this version but they should be added if there is a need for a more holistic view of the system.

In future versions of this database, an extra time-series table will be added considering the tilt and orientation of panels with tracking systems. The additional many-to-many table should be added to reflect the physical components, such as links between PV modules and battery storage when a direct (DC) link exists between PV and batteries.

\section{Conclusions}

The development of a scalable relational database for hosting data from PV research has been proposed. The relationships between the tables follow the physical connections of the modules while each system part has its own identity for historical tracking reasons (equipment movement etc.). Every measurement should be connected to all physical equipment before it, as well as to the operator of the specific site. It has been shown, that on time-series data, the proper usage of the Ramer-Douglas-Peucker algorithm along with Timescale DB compression, can save more than 98\% of disk space while increasing the performance of queries.
This database can be used for modeling, comparative analysis, O\&M systems as well as other applications. 

{
\footnotesize%
\bibliographystyle{IEEEtran}
\bibliography{bibliography}

\begin{thebibliography}{10}
\providecommand{\url}[1]{#1}
\csname url@samestyle\endcsname
\providecommand{\newblock}{\relax}
\providecommand{\bibinfo}[2]{#2}
\providecommand{\BIBentrySTDinterwordspacing}{\spaceskip=0pt\relax}
\providecommand{\BIBentryALTinterwordstretchfactor}{4}
\providecommand{\BIBentryALTinterwordspacing}{\spaceskip=\fontdimen2\font plus
\BIBentryALTinterwordstretchfactor\fontdimen3\font minus
  \fontdimen4\font\relax}
\providecommand{\BIBforeignlanguage}[2]{{%
\expandafter\ifx\csname l@#1\endcsname\relax
\typeout{** WARNING: IEEEtran.bst: No hyphenation pattern has been}%
\typeout{** loaded for the language `#1'. Using the pattern for}%
\typeout{** the default language instead.}%
\else
\language=\csname l@#1\endcsname
\fi
#2}}
\providecommand{\BIBdecl}{\relax}
\BIBdecl

\bibitem{ELAMIM2018121}
\BIBentryALTinterwordspacing
A.~Elamim, B.~Hartiti, A.~Haibaoui, A.~Lfakir, and P.~Thevenin, ``Performance
  evaluation and economical analysis of three photovoltaic systems installed in
  an institutional building in errachidia, morocco,'' \emph{Energy Procedia},
  vol. 147, pp. 121--129, 2018, international Scientific Conference
  “Environmental and Climate Technologies”, CONECT 2018, 16-18 May 2018,
  Riga, Latvia. [Online]. Available:
  \url{https://www.sciencedirect.com/science/article/pii/S1876610218301978}
\BIBentrySTDinterwordspacing

\bibitem{Arafet}
K.~Arafet and R.~Berlanga-Llavori, ``Digital twins in solar farms: An approach
  through time series and deep learning,'' \emph{Algorithms}, vol.~14, p. 156,
  05 2021.

\bibitem{Naik}
R.~Naik, A.~Tiihonen, J.~Thapa, C.~Batali, Z.~Liu, S.~Sun, and T.~Buonassisi,
  ``Discovering equations that govern experimental materials stability under
  environmental stress using scientific machine learning,'' \emph{npj
  Computational Materials}, vol.~8, p.~72, 04 2022.

\bibitem{4Mujumdar}
U.~B. Mujumdar and D.~R. Tutkane, ``Development of integrated hardware set up
  for solar photovoltaic system monitoring,'' in \emph{2013 Annual IEEE India
  Conference (INDICON)}, 2013, pp. 1--6.

\bibitem{5Touati}
F.~Touati, M.~Al-Hitmi, N.~Chowdhury, J.~Hamad, and A.~J. Gonzales,
  ``Investigation of solar pv performance under doha weather using a customized
  measurement and monitoring system,'' \emph{Renewable Energy}, vol.~89, pp.
  564--577, 04 2016.

\bibitem{6GAD2015337}
\BIBentryALTinterwordspacing
H.~Gad and H.~E. Gad, ``Development of a new temperature data acquisition
  system for solar energy applications,'' \emph{Renewable Energy}, vol.~74, pp.
  337--343, 2015. [Online]. Available:
  \url{https://www.sciencedirect.com/science/article/pii/S0960148114004649}
\BIBentrySTDinterwordspacing

\bibitem{7HERTELEER2017408}
\BIBentryALTinterwordspacing
B.~Herteleer, B.~Huyck, F.~Catthoor, J.~Driesen, and J.~Cappelle, ``Normalised
  efficiency of photovoltaic systems: Going beyond the performance ratio,''
  \emph{Solar Energy}, vol. 157, pp. 408--418, 2017. [Online]. Available:
  \url{https://www.sciencedirect.com/science/article/pii/S0038092X1730717X}
\BIBentrySTDinterwordspacing

\bibitem{8Zdanowicz}
T.~Zdanowicz, M.~Prorok, W.~Kolodenny, and H.~Roguszczak, ``Outdoor data
  acquisition system with advanced database for pv modules characterization,''
  06 2003, pp. 2497 -- 2500 Vol.3.

\bibitem{9Meliones}
A.~Meliones and A.~Nouvaki, ``A web-based three-tier control and monitoring
  application for integrated facility management of photovoltaic systems,''
  \emph{Applied Computing and Informatics}, vol.~10, 01 2014.

\bibitem{10Nihar}
A.~Nihar, A.~J. Curran, A.~M. Karimi, J.~L. Braid, L.~S. Bruckman,
  M.~Koyutürk, Y.~Wu, and R.~H. French, ``Toward findable, accessible,
  interoperable and reusable (fair) photovoltaic system time series data,'' in
  \emph{2021 IEEE 48th Photovoltaic Specialists Conference (PVSC)}, 2021, pp.
  1701--1706.

\bibitem{11Percuku}
A.~Perçuku, D.~Minkovska, L.~Stoyanova, and A.~Abdullahu, ``Iot using
  raspberry pi and apache cassandra on pv solar system,'' 09 2020, pp. 1--5.

\bibitem{12RAMER1972244}
\BIBentryALTinterwordspacing
U.~Ramer, ``An iterative procedure for the polygonal approximation of plane
  curves,'' \emph{Computer Graphics and Image Processing}, vol.~1, no.~3, pp.
  244--256, 1972. [Online]. Available:
  \url{https://www.sciencedirect.com/science/article/pii/S0146664X72800170}
\BIBentrySTDinterwordspacing

\bibitem{13Douglas1973ALGORITHMSFT}
D.~H. Douglas and T.~K. Peucker, ``Algorithms for the reduction of the number
  of points required to represent a digitized line or its caricature,''
  \emph{Cartographica: The International Journal for Geographic Information and
  Geovisualization}, vol.~10, pp. 112--122, 1973.

\end{thebibliography}
}
\end{document}